\documentclass[aps,
prb,
reprint,
superscriptaddress,
footinbib,
amsfonts,
amssymb,
amsmath,
intlimits,
]{revtex4-2}

\usepackage{bm,latexsym,mathrsfs,enumerate,bbm}
\usepackage{upgreek}
\usepackage[table,x11names]{xcolor}
\usepackage[breaklinks=true,
unicode=true,
urlcolor = RoyalBlue4,
colorlinks = true,
citecolor = RoyalBlue4,
linkcolor = RoyalBlue4
]{hyperref}
\usepackage{graphicx}
\usepackage[labelformat=simple]{subcaption}

\usepackage{chemformula}
\usepackage{siunitx}
\usepackage{enumitem}
\usepackage{mathtools}
\usepackage[cal=boondoxo,scr=rsfs]{mathalfa}
\renewcommand{\vec}[1]{\bm{#1}}
\begin{document}

\title{Magnetic Vortex Dynamics on a Spherical Cap}

\author{Mykola I. Sloika}
\email{sloika.m@gmail.com}
\affiliation{Taras Shevchenko National University of Kyiv, 01601 Kyiv, Ukraine}

\author{\boxed{\text{Yuri Gaididei}}}
\affiliation{Bogolyubov Institute for Theoretical Physics of the National Academy of Sciences of Ukraine, 03143 Kyiv, Ukraine}

\author{Volodymyr P. Kravchuk}
\email{v.kravchuk@ifw-dresden.de}
\affiliation{Leibniz-Institut für Festkörper- und Werkstoffforschung, IFW Dresden, 01171 Dresden, Germany}
\affiliation{Bogolyubov Institute for Theoretical Physics of the National Academy of Sciences of Ukraine, 03143 Kyiv, Ukraine}

\author{Oleksandr~V.~Pylypovskyi}
\email{o.pylypovskyi@hzdr.de}
\affiliation{Helmholtz-Zentrum Dresden-Rossendorf e.V., Institute of Ion Beam Physics and Materials Research, 01328 Dresden, Germany}
\affiliation{Kyiv Academic University, 03142 Kyiv, Ukraine}

\author{Denys Makarov}
\email{d.makarov@hzdr.de}
\affiliation{Helmholtz-Zentrum Dresden-Rossendorf e.V., Institute of Ion Beam Physics and Materials Research, 01328 Dresden, Germany}

\author{Denis D. Sheka}
\email{sheka@knu.ua}
\affiliation{Taras Shevchenko National University of Kyiv, 01601 Kyiv, Ukraine}

\date{November 15, 2024}

%%%%%%%%%%%%%%%%%%%%%%%%%%%%%%%%%%%%%%%%%%%%%%%%%%%%%%%%%%%%%%%%%%%%%70
%
%         ABSTRACT
%
%%%%%%%%%%%%%%%%%%%%%%%%%%%%%%%%%%%%%%%%%%%%%%%%%%%%%%%%%%%%%%%%%%%%%70

\begin{abstract}
By tailoring geometrical properties of magnetic nanocap structures, new possibilities appear to control its magnetic properties, such as the dynamics of magnetic vortices. Here, we develop an approach to describe dynamics of magnetic vortices on a spherical cap. Analytic results for the gyrofrequency of the vortex state are in a good agreement with micromagnetic simulations.
\end{abstract}


% 75.75.-c	  Magnetic properties of nanostructures
% 75.78.-n	  Magnetization dynamics
% 75.78.Jp    Ultrafast magnetization dynamics and switching
% 75.78.Cd    Micromagnetic simulations

%%%%%%%%%%%%%%%%%%%%%%%%%%%%%%%%%%%%%%%%%%%%%%%%%%%%%%%%%%%%%%%%%%%%%70

\maketitle

%%%%%%%%%%%%%%%%%%%%%%%%%%%%%%%%%%%%%%%%%%%%%%%%%%
%%%			Introduction
%%%%%%%%%%%%%%%%%%%%%%%%%%%%%%%%%%%%%%%%%%%%%%%%%%

\section{Introduction}

Curvilinear magnetism is a framework to analyze geometry-governed effects and predict modifications of magnetic responses in geometrically curved nanostructures \cite{Makarov22a}. The theory introduces emergent interactions, which are are responsible for different families of curvature-induced effects including topological patterning and magnetochiral effects \cite{Vedmedenko20}. For instance, the appearance of a skyrmion state in core-shell particles with easy-normal anisotropy \cite{Kravchuk16a} is a fingerprint of the topological patterning. Other examples include nonorientable  Möbius rings, which support magnetic vortex states and different topologically protected domain walls \cite{Pylypovskyi15b}, and spherical shells with easy-surface anisotropy favouring a 3D onion state in thin spherical shells \cite{Kravchuk12a} and whirligig state in thick shells \cite{Sloika17}. 

Among different curvilinear magnetic systems, there is one group of the so-called spherical cap structures, which can be prepared by deposition of a magnetic thin film on non-magnetic spheres \cite{Streubel16a, Makarov22a}. Depending on the magnetic anisotropy of the thin film, either radially magnetized magnetic structures or magnetic vortices can be realized in magnetic nanocaps. Magnetic vortices were actively studied in planar films and planar nanodots \cite{Hubert09}. A vortex state is characterized by the in-plane magnetization circulation following the edge of a nanodot and possessing an out-of-plane magnetization component (polarity) at the location of the vortex core. In the case of curved geometries, magnetic vortices acquire new properties in comparison with their planar counterparts. This includes polarity-circulation coupling of the vortex on a spherical cap and shell \cite{Kravchuk12a, Sheka22b}, controllable switching of the vortex magnetochirality on a hemispherical cap \cite{Yershov15}, chirality symmetry breaking upon switching of the vortex core \cite{Sloika14}, and nonlocal chiral symmetry breaking effect in cap structures \cite{Sheka20a, Volkov23}. Spherical geometries are well studied with respect to static magnetic distributions \cite{Streubel12a,Streubel12,Kravchuk12a, Sheka13a,Sloika17,Sloika22}. 

The study of magnetization dynamics of topological textures is one of the important
and challenging problems of the curvilinear magnetism. At the moment, dynamics of topological soliton is well described mainly for the tubular geometry \cite{Landeros22}. General aspects of the dynamics of solitons in curved geometries are studied for skyrmions \cite{Korniienko20, Sheka22b}. In particular, the skyrmion gyromotion was studied analytically using a Thiele-like collective variable approach \cite{Korniienko20}. Investigations of magnetic dynamics of vortices in curved geometries include the description of resonance excitations \cite{Kim15a}, polarity and circulation switching \cite{Streubel12a,Sloika14,Yershov15,Streubel16a}. The use of the Thiele-like approach for the vortex dynamics becomes more complex due to their nonlocalized nature. 

Here, we develop a theoretical approach to describe the motion of magnetic vortices in nanoshells of spherical geometry. The method is based on the image vortex technique~\cite{Kovalev03a,Sheka05}, which we extend to curved geometries. Using mapping from a sphere to a complex plane, we derive equations of motion of magnetization on a spherical cap. We calculate the gyrofrequency of the vortex core and validate our analytic theory with micromagnetic simulations. 

%%%%%%%%%%%%%%%%%%%%%%%%%%%%%%%%%%%%%%%%%%%%%%%%%%
%%%	New Section
%%%%%%%%%%%%%%%%%%%%%%%%%%%%%%%%%%%%%%%%%%%%%%%%%%

\section{Magnetic vortices on a spherical cap}

\begin{figure}
	\includegraphics[width=\columnwidth]{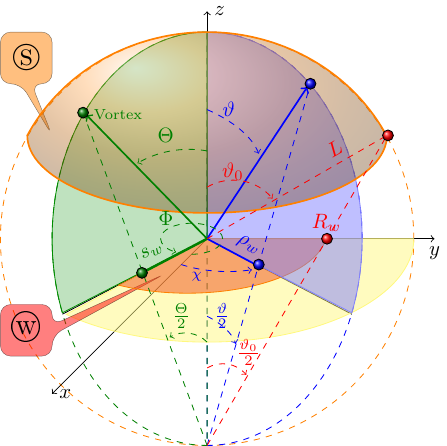}
	\caption{(Colour online) \textbf{Mapping schematic $\textcircled{s}\mapsto \textcircled{w}$:} Stereographic projection of a cap (part of a sphere, $\circledS$) to a complex plane $\textcircled{w}$.}
	\label{fig:mapping}
\end{figure}

We consider a spherical cap of the inner radius $L$, thickness $h$ and the cut angle $\vartheta_0$,  see Fig.~\ref{fig:mapping}. The curved geometry brings about two geometry-governed emergent magnetic interactions: effective anisotropy and effective Dzyaloshinskii–Moriya interaction (DMI) \cite{Gaididei14}. Depending on the geometrical and material parameters, these emergent interactions might significantly change magnetic vortex properties. Namely, similar to a spherical shell \cite{Kravchuk16a},the effective DMI leads to an additional angular correction for the in-surface magnetic distribution. In the case of a spherical cap this leads to (i) the magnetization being not in the surface at the cap border and (ii) an increase of the surface energy. At the same time, the effective DMI slightly increases the size of the magnetic vortex core for any combination of the vortex polarity and circulation~\cite{Sloika22}. These effects are considerable for thin spherical caps with a small inner radius $L$.

In the current study, we describe a thin spherical cap with a large inner radius, supposing that $h\ll\ell\ll L$, where $\ell=\sqrt{A/(2\pi M_s^2)}$ is the exchange length with $A$ being the exchange constant and $M_s$ being the saturation magnetization. Under this assumption, the magnetization does not depend on the radial coordinate $r$ and we can neglect curvature induced effects. 
It is convenient to use the local spherical reference frame for the unit magnetization vector $\vec{m} = (m_r, m_\vartheta, m_\chi) = (\cos\theta, \sin\theta\cos\phi, \sin\theta\sin\phi)$. Here, angular magnetic variables $\theta=\theta(\vec r)$ and $\phi=\phi(\vec r)$ describe the magnetization distribution with respect to the spherical coordinates $(r,\vartheta,\chi)$ of the radius--vector $\vec{r}$.

For the case of an isotropic magnet, the energy $E$, normalized by $E_0=2\pi M_s^2 V$, can be presented as $\mathscr{E} = E/E_0=\mathscr{E}_{\text{x}} + \mathscr{E}_{\text{ms}}$. The exchange energy
$\mathscr{E}_{\text{x}} =- (\ell^2/V) \int_V\!\! \mathrm{d}\vec{r} \left( \vec{m}\cdot\vec{\nabla}^2\vec{m} \right)$. The second term is the magnetostatic energy, which can be represented via magnetostatic volume charges, $\lambda_{\text{ms}} = -\vec{\nabla}\cdot \vec{m}$, and surface charges, $\sigma_{\text{ms}} = \vec{m}\cdot \vec{n}$ with $\vec{n}$ being the external normal to the surface. The magnetostatic potential $\Phi_{\text{ms}}(\vec{r})$ and the corresponding energy $\mathscr{E}_{\text{ms}}$ read:
\begin{equation} \label{eq:E-ms}
\begin{aligned}
\Phi_{\text{ms}}(\vec{r}) &= \int\limits_V\!\! \mathrm{d}\vec{r}' \frac{\lambda_{\text{ms}}(\vec{r}')}{|\vec{r} - \vec{r}'|} + \int\limits_S \mathrm{d}S' \frac{\sigma_{\text{ms}}(\vec{r}')}{|\vec{r} - \vec{r}'|},\\
\mathscr{E}_{\text{ms}} = \frac{1}{8\pi V} \Biggl[&\int\limits_V \mathrm{d}\vec{r} \lambda_{\text{ms}}(\vec{r})\Phi_{\text{ms}}(\vec{r})\\
 &+ \int\limits_S \mathrm{d}S \sigma_{\text{ms}}(\vec{r})\Phi_{\text{ms}}(\vec{r})\Biggr].
\end{aligned}
\end{equation}
The ground state of the spherical cap is known to depend on the material and geometrical parameters of the sample \cite{Sheka13b, Streubel12}. In particular, depending on the cap radius and thicknesses, different magnetization textures can be realized: (i) the monodomain (onion) state is favourable for thin samples, (ii) the uniform easy--axis state appears for a thick cap of a small diameter, and (iii) the vortex state appears for relatively large samples \cite{Sheka13b,Streubel12}. To describe the geometrical parameters, it is convenient to introduce the aspect ratio of a spherical cap as follows:
\begin{equation} \label{eq:aspect-ratio}
\begin{split}
\varepsilon &= \frac{\text{edge area}}{\text{surface area (in+out)}}\\
&= \frac{\nu\cot(\vartheta_0/2)}{2}\cdot \frac{1+\nu/2}{1+\nu+\nu^2/2}, \quad \nu = \frac{h}{L}.
\end{split}
\end{equation}
We note that this definition of the aspect ratio corresponds to the standard one of a disk--shaped sample, $\varepsilon=h^{\text{disk}}/L^{\text{disk}}$ in the limiting case $\vartheta_0\to0$, $L\to\infty$ under the constraint $L\vartheta_0\to L^{\text{disk}}=\text{const}$.

For a spherical cap with the equilibrium magnetization distribution in the form of an out--of--surface vortex, the vortex at the northern pole of the cap can be described by the following Ansatz \cite{Kravchuk12a}:
\begin{equation} \label{eq:vortex}
\cos\theta(\vartheta) = p\,m(\vartheta), \quad \phi(\chi) = \mathfrak{C}\frac\pi2.
\end{equation}
Here $p=\pm1$ is the vortex polarity (upward or downward) and  $\mathfrak{C}=\pm1$ is the vortex circulation (clockwise or counterclockwise). The shape of the out--of--surface vortex structure is described by the function $m(\vartheta)$, which is exponentially localized at $\vartheta=0$ with the limiting values: $m(0)=p$ and $m(\vartheta_0)=0$ similar to the out--of--surface vortex in a planar disk sample \cite{Feldtkeller65}.

%==================================================================\
%==================================================================/

%%%%%%%%%%%%%%%%%%%%%%%%%%%%%%%%%%%%%%%%%%%%%%%%%%
%%%	New Section
%%%%%%%%%%%%%%%%%%%%%%%%%%%%%%%%%%%%%%%%%%%%%%%%%%

\section{Gyroscopic vortex dynamics on a spherical cap}

We describe the dynamics of a vortex relying on a collective variable method similar to the established approaches for planar magnets. For this purpose, we map a spherical surface to a complex $z$--plane. It is instructive to visualize the mapping as follows: (i) Using the stereographic projection, we map part of a sphere to a complex $w$--plane: $\circledS\mapsto \textcircled{w}$, see Fig.~\ref{fig:mapping}. Under such mapping, $\rho_w =L\tan\frac{\vartheta}{2}$. (ii) We map the circle in the $w$--plane to the unit circle in the $z$--plane: $\textcircled{w}\mapsto \textcircled{z}$, which is a scaling transformation, $\rho=\rho_w/(LR)$. Finally
\begin{equation} \label{eq:mapping}
\rho =\frac{1}{R}\tan\frac{\vartheta}{2},\qquad \rho\in[0,1), \qquad R = \tan\frac{\vartheta_0}{2}.
\end{equation}

\begin{figure}
	\includegraphics[width=0.9\columnwidth]{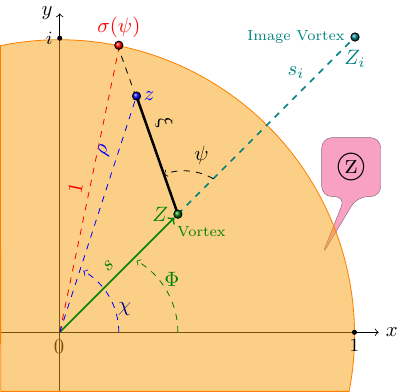}
	\caption{(Colour online) \textbf{Local reference frame in \textcircled{z}-plane:} $\xi = |z-Z|$ and $\psi = \arg(z-Z)-\arg{Z}$. Here $z$ is a representation point on a complex plane, $Z$ and $Z_i$ are the vortex position and the position of its image on the plane. }
	\label{fig:mobile}
\end{figure}

Here, we use a complex variable $z=\rho e^{i\chi}$ to describe the mapping of the radius--vector $\vec{r}$ to the complex plane. After mapping of a spherical cap $\circledS$ to the complex plane \textcircled{z}, we can use methods known from the studying the vortex dynamics in a planar geometry. The simplest way is to use the Thiele approach \cite{Thiele73,Thiele73}, where the vortex moves like a rigid particle without changing its shape: $\vec{m}(z,t)=\vec{m}\left(z-Z(t)\right)$ with $Z(t)$ being the position of the vortex centre of mass. However, such approach does not take into account the boundary effects, which are important for a nanoparticle. In particular, it is well--known \cite{Guslienko02a} that the rigid vortex model does not provide the correct gyroscopic frequency of the vortex motion. To describe the boundary effects, we use the \emph{image vortex ansatz} \cite{Mertens94a} in the following form, see Fig.~\ref{fig:mobile}:
\begin{equation} \label{eq:iva}
\begin{split}
&\cos \theta^{\textsc{iva}}(z,t) = p \mu(|z-Z(t)|),\\
&\phi^{\textsc{iva}}(z,t)   = \arg(z-Z(t))+\arg(z-Z_i(t))\\
& -\arg Z(t)-\arg z + \mathfrak{C}\frac\pi2,\\
&Z(t)   = s(t)   e^{i\Phi(t)}, \quad s(t)   = \frac{1}{R}\tan\frac{\Theta(t)}{2}\in[0,1)\\
&Z_i(t) = s_i(t) e^{i\Phi(t)}, \quad s_i(t) = \frac{1}{s(t)},\\
&\mu(\rho) \approx \exp\left(-\frac{\rho^2}{2\lambda^2} \right), \qquad \lambda = \frac{\ell}{RL}\ll1,
\end{split}
\end{equation}
where $\mu(\rho) \equiv m(\vartheta)$ and $\lambda\ll1$ is the dimensionless vortex core size. For simplicity, we consider $\cos \theta_0 = p$. 

The stereographic projection used here possesses two important features: (i) being a conformal mapping, it keeps a Laplace equation invariant, and (ii) it maps \emph{finite size} circles on a sphere to finite size circles on a plane~\cite{Tristan21}. For the magnetization distribution tangential to the sphere surface ($\theta\equiv\pi/2$), the minimizer of the exchange energy $\mathscr{E}_{\text{x}} = (\ell^2/V) \int_V[(\vec{\nabla}\phi-\vec{\Omega})^2+r^{-2}]\mathrm{d}\vec{r}$~\cite{Gaididei14} is a function $\phi(\vartheta,\chi)$, which satisfies the Laplace equation $\vec{\nabla}^2\phi=0$. Here $(\Omega_r,\Omega_\vartheta, \Omega_\chi)= (0,0,-\cot\vartheta/r)$ is the vector of the spin connection. Note that $\vec{\nabla}^2\phi_{\textsc{iva}}=0$, which reflects the invariance of the Laplace equation under a conformal mapping. Thus, the usage of the Ansatz \eqref{eq:iva} for $\phi^{\textsc{iva}}$ corresponds to the exchange approximation in which the magnetostatic contribution is approximated by the easy-surface anisotropy. However, the vortex gyration is mostly determined by the energy of the volume magnetostatic charges~\cite{Guslienko02a,Ivanov07b,Gaididei10}. In what follows, the latter contribution is treated as the first-order perturbation. Finally, the feature (ii) of the stereographic projection enables us to model the out-of-surface component $\cos \theta^{\textsc{iva}}$ by a radially-symmetric function.

%==================================================================\
%==================================================================/

\begin{figure}	
	\includegraphics [width=1.\columnwidth,angle=0]{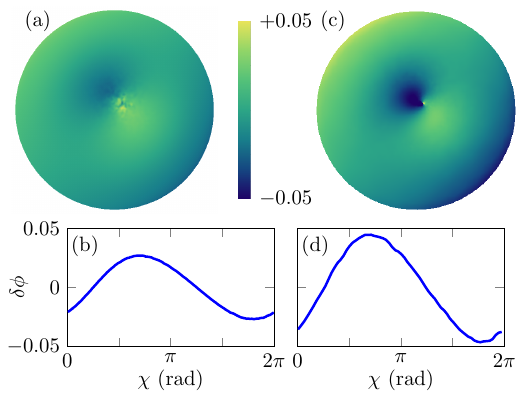}
	\caption{(Colour online) \textbf{Deviation of the image vortex model from the real magnetic distribution:} The difference between the simulation data and the image vortex model for a disk (a,b) and a cap (c,d). Colour density plots (a,c) correspond to the difference between the calculated from simulation data $\phi$ and $\phi^{\textsc{iva}}$. Plots (b,d) correspond to the deviation on the border of the nanoparticle ($\rho=1$). The vortex displacement for both geometries is the same ($s=0.1$) as well as the aspect ratio ($\varepsilon=0.48$).}
	\label{fig:deltaPhiVsChi}
\end{figure}

To verify that the mapping of a sphere on a complex plane and the usage of the image vortex model provide an adequate magnetization texture, we use micromagnetic simulations.  For the simulations, we use \textsf{magpar} code~\cite{MAGPAR, Scholz03a} with Permalloy material parameters \footnote{Permalloy is chosen as a material with the following parameters: the exchange constant $A = 21$~pJ/m, the saturation magnetization $M_s = 795$~kA/m and damping constant $\alpha=0.01$. These parameters result in the exchange length $\ell\approx 5.14$~nm. Thermal effects and anisotropy are neglected.}. Namely, we shift the vortex from the centre of a nanoparticle of hemispherical cap and planar disk geometries. Next we map the magnetization texture of a spherical cap, obtained from simulations, to a complex plane and determine the vortex centre as intersection of isolines $m_\vartheta = 0$ and $m_\chi = 0$. Then we extract the magnetic angle $\phi$ from the simulation and compare it with the corresponding angle $\phi^{\textsc{iva}}$ obtained from the image vortex model. In~Fig.~\ref{fig:deltaPhiVsChi} we show that the deviation of the image vortex model from the results of micromagnetic simulations $\delta\phi=\phi-\phi^{\textsc{iva}}$ in the case of a hemispherical cap is mostly the same as in the case of a disk.

%==================================================================\
%==================================================================/

The magnetization dynamics is governed by the Landau--Lifshitz--Gilbert equations, which can be derived from the D\"oring Lagrangian and Rayleigh dissipative function:
\begin{align} \label{eq:L}
\mathscr{L}[\theta,\phi] &= \frac1V\int_V\!\! \mathrm{d}\vec{r} (\cos\theta_0-\cos\theta)\dot {\phi} - \mathscr{E}[\theta,\phi],\\
\label{eq:F} %
\mathscr{F}[\theta,\phi] &= \frac{\alpha}V\int_V\!\! \mathrm{d}\vec{r} \left( {\dot{\theta}}^2 + \sin^2\theta {\dot{\phi}}^2\right).
\end{align}
Here the overdot denotes the derivative with respect to the rescaled time in units of $\omega_0^{-1}=(4\pi M_S\gamma)^{-1}$ with $\gamma$ being the gyromagnetic ratio, $\mathscr{E}$ is the system energy, normalized by the value $4\pi M_S^2V$, $V$ is the sample volume, $\alpha$ is the dissipative constant, and $\theta_0$ is an arbitrary constant.

Now we can derive the effective Lagrangian by incorporating Ansatz \eqref{eq:iva} in the \eqref{eq:L}:
\begin{equation} \label{eq:L-eff}
\begin{split}
L &= G - E,\qquad E = \mathscr{E}\left[\theta^{\textsc{iva}}, \phi^{\textsc{iva}}\right],\\
G &= \frac{1}{V}\int_V\!\! \mathrm{d}\vec{r} \left(p - \cos\theta^{\textsc{iva}} \right) \dot{\phi}^{\textsc{iva}},
\end{split}
\end{equation}
Here we regularized the effective Lagrangian \eqref{eq:L} by choosing $\cos\theta_0=p$~\cite{Sheka06e}. The direct calculation results in the following effective gyroscopic term (see Appendix \ref{sec:Gyro} for details)
\begin{equation} \label{eq:G-eff}
G = \frac{1+R^2}{1+R^2s^2}\,p\, s^2\dot\Phi  = \frac{1-\cos\Theta}{1-\cos\vartheta_0}\,p\,\dot\Phi.
\end{equation}
The curvature results in a nonlinear effect in the gyroscopic term. In comparison, in the limiting case of a planar magnet ($R\ll1$), the gyroscopic term $G^{\text{disk}} = p s^2 \dot{\Phi}$, see e.g. Ref.~\onlinecite{Sheka05}.

In the same way, we derive the effective dissipative function using the Ansatz \eqref{eq:iva}. For a large sample, $\ln\left(LR/\ell\right)\gg1$, we calculate the dissipative function $F = \mathscr{F}\left[\theta^{\textsc{iva}}, \phi^{\textsc{iva}}\right]$ as follows (see Appendix \ref{sec:Fdiss} for details):
\begin{equation} \label{eq:F-eff}
\begin{split}
F &= \frac{\eta\left(1+R^2\right)}{\left(1+R^2 s^2\right)^2}\left( \dot{s}^2 + s^2\dot{\Phi}^2\right)\\
&= \frac{\eta}{2\left(1-\cos\vartheta_0\right)}\left(\dot{\Theta}^2+\sin^2\Theta\dot{\Phi}^2 \right),\\
\eta &= \alpha \ln\frac{1}{\lambda} = \alpha\ln\frac{RL}{\ell} = \alpha\ln\frac{L\tan(\vartheta_0/2)}{\ell}.
\end{split}
\end{equation}
We rescale the effective Lagrangian $L^{\text{ef}} = (1-\cos\vartheta_0)L$, the dissipative function $F^{\text{ef}} = (1-\cos\vartheta_0)F$, and the effective energy $E^{\text{ef}} = (1-\cos\vartheta_0)E$:
\begin{equation} \label{eq:L&F-ef}
\begin{split}
L^{\text{ef}} &= (1-\cos\Theta) p \dot{\Phi} - E^{\text{ef}}[\Theta,\Phi],\\
F^{\text{ef}} &= \frac{\eta}{2}\left( \dot{\Theta}^2 + \sin^2\Theta \dot{\Phi}^2\right).
\end{split}
\end{equation}

Assuming that $\dot{p}=0$, we write the corresponding Euler--Lagrange--Rayleigh equations 
\begin{equation} \label{eq:ELR-eff}
\begin{split}
p  \sin\Theta \dot{\Phi}  &= \frac{\partial E^{\text{ef}}}{\partial \Theta} + \eta \dot{\Theta},\\
-p \sin\Theta \dot{\Theta}  &= \frac{\partial E^{\text{ef}}}{\partial \Phi} + \eta \sin^2\Theta \dot{\Phi}.
\end{split}
\end{equation}

It is instructive to draw an intermediate conclusion concerning the form of the effective equations of the vortex motion. For the case $p=\pm1$ and by introducing the variable $\cos\Psi =  p\cos\Theta$, the Euler--Lagrange--Rayleigh Eqs.~\eqref{eq:ELR-eff} can be rewritten as follows:
\begin{equation} \label{eq:ELR-Psi-Phi}
\begin{split}
\sin\Psi \dot{\Phi}   &= \frac{\partial E^{\text{ef}}}{\partial \Psi} + \eta \dot{\Psi},\\
-\sin\Psi \dot{\Psi}  &= \frac{\partial E^{\text{ef}}}{\partial \Phi} + \eta \sin^2\Psi \dot{\Phi}.
\end{split}
\end{equation}
The Eqs.~\eqref{eq:ELR-Psi-Phi} have formally the form of the Landau--Lifshitz--Gilbert equation for angular variables. The reason is that the ``vortex centre coordinate'' $\vec{X} = \left(\sin\Psi \cos\Phi, \sin\Psi \sin\Phi, \cos\Psi\right)$ is a unit vector, which gyrates on a sphere in the same way as the magnetization vector evolves following the Landau--Lifshitz--Gilbert equation:
\begin{equation} \label{eq:X-LLE}
\dot{\vec{X}} = \vec{X}\times \frac{\partial E^{\text{ef}}}{\partial \vec{X}} + \eta \vec{X}\times \dot{\vec{X}}.
\end{equation}
We rewrite Eq.~\eqref{eq:X-LLE} in the Thiele--like form
\begin{equation} \label{eq:X-Thiele}
\begin{split}
&\vec{F}^{\text{gyro}}+\vec{F}^{\text{ext}}+\vec{F}^{\text{dis}}=0, \\
&\vec{F}^{\text{gyro}} = \dot{\vec{X}}\times \vec{X}, \; \vec{F}^{\text{ext}} = - \frac{\partial E^{\text{ef}}}{\partial \vec{X}}, \; \vec{F}^{\text{dis}} = -\eta\dot{\vec{X}}.
\end{split}
\end{equation}
The meaning of the last equation is the force balance condition between the gyroscopic force $\vec{F}^{\text{gyro}}$, the external force $\vec{F}^{\text{ext}}$, and the dissipative force $\vec{F}^{\text{dis}}$.

%%%%%%%%%%%%%%%%%%%%%%%%%%%%%%%%%%%%%%%%%%%%%%%%%%
%%%	New Section
%%%%%%%%%%%%%%%%%%%%%%%%%%%%%%%%%%%%%%%%%%%%%%%%%%

\section{Magnetostatic energy and gyrofrequency}

We calculate the energy of a moving vortex in a spherical cap. Similar to the vortex gyroscopic motion in a planar disk \cite{Guslienko02a,Ivanov07b,Gaididei10}, it is expected that the main contribution to the effective energy is due to the stray field energy of the shifted vortex. More precisely, the origin of the stray field is the volume magnetostatic charges, $\lambda_{\text{ms}}$. By neglecting the out--of--surface contribution of the vortex structure and exploring the image vortex Ansatz \eqref{eq:iva}, we calculate the magnetostatic vortex energy as follows:
\begin{equation} \label{eq:Ems-via-C}
\begin{split}
\mathscr{E}_{\text{ms}} = \varepsilon \frac{1-\cos\Theta}{1-\cos\vartheta_0}\mathcal{C}(\nu,s,\vartheta_0),
\end{split}
\end{equation}
where $\mathcal{C}(\nu,s,\vartheta_0)$ is a function that can be derived for small vortex displacements $s \ll 1$ (see Appendix \ref{sec:Magnetostatic-app} for details).

For a very thin cap, $\nu\ll1$, and assuming that $|\sin\vartheta_0|$ is not very small (which means that the geometry is far from the planar case and full spherical shell), we can consider that $f_l(\nu\to0)=1$. Hence, the function $\mathcal{C}(\vartheta_0)\equiv \mathcal{C}(0,0,\vartheta_0)$ reads
\begin{equation} \label{eq:C4nu0}
\mathcal{C}(\vartheta_0) = \frac{1}{2\sin\vartheta_0 \left(1-\cos\vartheta_0\right)}  \sum_{l=1}^\infty \frac{I_l^2(\vartheta_0)}{l(l+1)}.
\end{equation}
For an approximate description of the behaviour \eqref{eq:C4nu0}, we use the trial function
\begin{equation} \label{eq:C4nu0-fit}
\mathcal{C}^{\text{fit}}(\vartheta_0) = \frac16 +\frac1{13}\exp\left(-1.7\cos\vartheta_0\right).
\end{equation}

Now we consider the case $R\ll1$, which corresponds to a planar magnet. We note that the series \eqref{eq:Ems&C} does not uniformly converge for $R\to0$. It is convenient to start from the stiffness parameter in the form \eqref{eq:C}. In this case, we calculate that $\mathcal{C}(\nu\sim2\varepsilon R,R,0)$ tends to the limiting value $\mathcal{C}^{\text{disk}} = \frac{2}{3\pi}(2\mathfrak{G}-1)\approx0.177$ \cite{Gaididei10} when $\varepsilon\to0$ and $R\to0$. The magnetostatic energy of a disc
$E^{\text{disc}}_{\text{ms}} = \mathcal{C}^{\text{disk}} \varepsilon s^2$.
For a full hemispherical cap ($R=1$), we find that $\mathcal{C}(R=1)\approx 0.247$.

Now we can derive the effective magnetostatic energy~\eqref{eq:L&F-ef} in the form $\mathscr{E}_{\text{ms}}^{\text{ef}} = \varepsilon  (1-\cos\Theta) \mathcal{C}(\nu,s,R)$.
The effective equations of motion \eqref{eq:ELR-eff} can be presented as follows
\begin{equation} \label{eq:eff-no-field}
\dot{\Phi} = \frac{\Omega_g}{1+\eta^2}, \qquad \dot{\Theta} = -\eta p \sin\Theta \dot{\Phi}.
\end{equation}
Without damping ($\eta=0$), Eqns.~\eqref{eq:eff-no-field} describe the gyroscopic motion with a constant frequency
\begin{equation} \label{eq:Omega_g}
\dot{\Phi} = p \Omega_g, \qquad \Omega_g  = \varepsilon \left[\mathcal{C} + \frac{R s}{2}\left(1+R^2s^2 \right)\partial_{s} \mathcal{C} \right].
\end{equation}
For the case of a very thin cap and small vortex displacements, we can use the following asymptotic expression for the gyrofrequency
\begin{equation} \label{eq:Omega_g0}
\Omega_g^0 =  \varepsilon \mathcal{C}(\vartheta_0),
\end{equation}
where $\mathcal{C}(\vartheta_0)$ is defined by \eqref{eq:C4nu0}. In the limiting case of a thin disk, $\Omega_g^{\text{disk}} = \varepsilon \mathcal{C}^{\text{disk}}\approx 0.177\varepsilon$ \cite{Gaididei10}.

%==================================================================\
\begin{figure}
	\includegraphics [width=1.\columnwidth,angle=0]{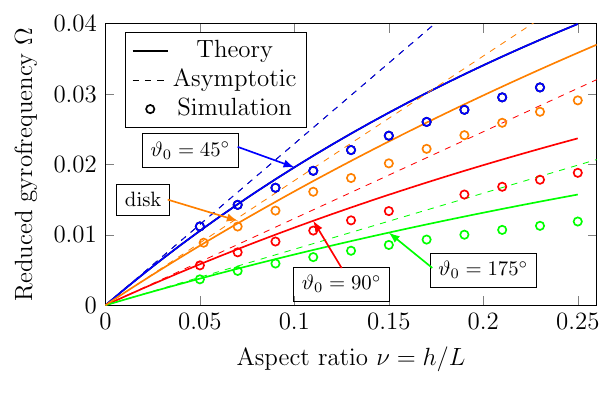}
	\caption{(Colour online) \textbf{Gyrofrequency vs. aspect ratio:} The vortex gyrofrequency depending on the nanoparticle aspect ratio $\nu=h/L$ for small vortex displacements. Solid lines correspond to calculated values of the gyrofrequency using (\ref{eq:Omega_g}). Dashed lines correspond to the asymptotic behaviour $\nu \ll 1$. Circles correspond to the gyrofrequency obtained from micromagnetic simulations. }
	\label{fig:omegaVsNu}
\end{figure}
%==================================================================/

\begin{figure}
	\includegraphics [width=1.\columnwidth,angle=0]{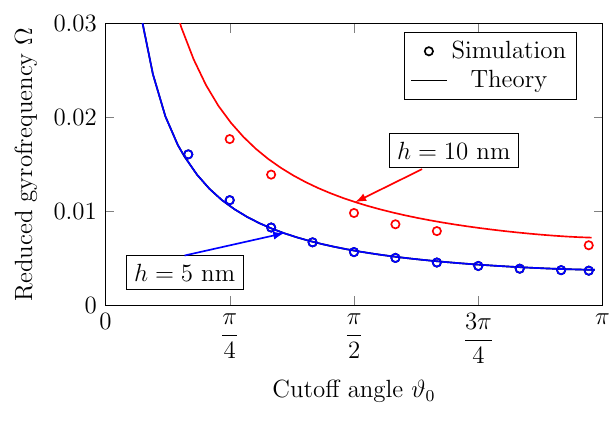}
	\caption{(Colour online) \textbf{Gyrofrequency vs. cutoff angle:} Circles correspond to the simulation results and the solid line is obtained from (\ref{eq:Omega_g}) for the case of a small vortex displacement. The inner radius of a spherical cap $L=100$~nm and thicknesses $h$ are specified in the plot.}
	\label{fig:omegaVsTheta}
\end{figure}

To verify our analytical estimations of the gyrofrequency, we perform a series of micromagnetic simulations in a wide range of geometric parameters. For simulations, we used the \textsf{magpar} code~\cite{MAGPAR, Scholz03a} with Permalloy material parameters \cite{Note1}. Numerically, we simulated 3D magnetization dynamics of the vortex on spherical caps with different aspect ratios and different cutoff angles. Initially, using a relaxation procedure, the magnetization was relaxed to the vortex state. Then, by applying a magnetic field pulse, the vortex was shifted from the centre followed by its motion along a spiral trajectory toward the origin. Using the Fourier analysis of magnetization oscillations, we calculated the vortex gyroscopic frequencies. The gyrofrequency, obtained from micromagnetic simulations, is displayed in Fig.~\ref{fig:omegaVsNu} and in Fig.~\ref{fig:omegaVsTheta} by circles for different aspect ratios and cutoff angles. We note that the obtained values of the gyrofrequency using Eq.~(\ref{eq:Omega_g}) are in a good agreement with the results of simulations in a wide range of aspect ratios and cutoff angles.

%==================================================================\
%==================================================================/

%%%%%%%%%%%%%%%%%%%%%%%%%%%%%%%%%%%%%%%%%%%%%%%%%%
%%%	New Section
%%%%%%%%%%%%%%%%%%%%%%%%%%%%%%%%%%%%%%%%%%%%%%%%%%

\section{Conclusion}

In conclusion, we developed an approach to describe the magnetic vortex motion on a spherical cap. The method is based on a conformal mapping of a spherical cap to a complex plane and usage of the image vortex ansatz. The method can be extended to describe dynamics of different magnetic textures in nanomagnets of a complex geometry. In the case of a spherical cap, this approach leads to the Thiele equation that has the form of the Landau-Lifshitz-Gilbert equation. This is explained by the fact that the ``vortex centre coordinate'' gyrates on a sphere in the same way as the magnetization vector. To demonstrate validity of the suggested approach, we calculated the gyrofrequency of a vortex core on spherical caps with different cutoff angles and aspect ratios. The theoretically obtained gyrofrequencies are in a good agreement with the data of micromagnetic simulations. Besides, the limiting case (inner radius $L\to\infty$ and cutoff angle $\vartheta_0\to0$) leads to the well-known solution for a planar disk.

%%%%%%%%%%%%%%%%%%%%%%%%%%%%%%%%%%%%%%%%%%%%%%%%%%
%%%			Acknowledgments
%%%%%%%%%%%%%%%%%%%%%%%%%%%%%%%%%%%%%%%%%%%%%%%%%%

\section*{Acknowledgments}
\label{sec:Acknowledgments}

The simulation results presented in the work were obtained using the computing cluster of the Taras Shevchenko National University of Kyiv \cite{unicc}. M.~S. and D.~S. acknowledge HZDR, where part of this work was performed, for hospitality. This work is financed in part via the German Research Foundation (DFG) under the Grant MC9/22-1, MA 5144/22-1, MA 5144/24-1, and ERC grant 3DmultiFerro (Project number: 101141331).

%%%%%%%%%%%%%%%%%%%%%%%%%%%%%%%%%%%%%%%%%%%%%%%%%%
%%%	Appendix
%%%%%%%%%%%%%%%%%%%%%%%%%%%%%%%%%%%%%%%%%%%%%%%%%%
\appendix

%%%%%%%%%%%%%%%%%%%%%%%%%%%%%%%%%%%%%%%%%%%%%%%%%%
%%%	New Section
%%%%%%%%%%%%%%%%%%%%%%%%%%%%%%%%%%%%%%%%%%%%%%%%%%

\section{Calculation of the effective gyroscopic term}
\label{sec:Gyro}

Here we calculate the effective gyroscopic term in the Lagrangian. For this purpose, we map a spherical surface coordinate to a complex plane: $\vec{r}\mapsto w\mapsto z$, see Fig.~\ref{fig:mapping}. The integration in these two systems are related by the following expression:
\begin{equation} \label{eq:dr}
\frac{1}{V}\int_V \!\! \mathrm{d}\vec{r} f(\rho,\chi)= \frac{1+R^2}{\pi} \!\! \int_0^{2\pi} \!\! \mathrm{d}\chi \!\! \int_0^1 \!\! \frac{f(\rho,\chi) \rho \mathrm{d}\rho}{\left(1+R^2\rho^2\right)^2} .
\end{equation}
Then the gyroscopic term $G = G_1 - G_2$ reads:
\begin{equation} \label{eq:G1&G2}
\begin{split}
G_1 &= \frac{p(1+R^2)}{\pi} \int_0^{2\pi} \mathrm{d}\chi \int_0^1 \frac{\dot{\phi}^{\textsc{iva}} \rho \mathrm{d}\rho}{\left(1+R^2 \rho^2\right)^2},\\
G_2 &= \frac{p(1+R^2)}{\pi} \int_0^{2\pi}\mathrm{d}\chi \int_0^1 \frac{\cos\theta^{\textsc{iva}} \dot{\phi}^{\textsc{iva}}\rho \mathrm{d}\rho}{\left(1+R^2 \rho^2\right)^2}.
\end{split}
\end{equation}

We derive $\dot\phi$ as follows:
\begin{equation} \label{eq:phi_t}
\begin{split}
& \dot{\phi}^{\textsc{iva}} =\phi_{\text{s}}\dot{s} -\phi_{\Phi}\dot{\Phi},\\
& \phi_{\text{s}} = \frac{\cos\psi}{\xi} -\dfrac{ \xi \sin \psi}{\xi^2 +(\rho^2-1)(s^2-1)},\\
& \phi_{\Phi} = \frac{s\sin\psi}{\xi} +\dfrac{ \xi s \cos\psi + s^2 -1}{\xi^2 +(\rho^2-1)(s^2-1)}+1,\\
& \xi=|z-Z(t)|,\qquad \psi = \arg\left(z-Z(t)\right)-\arg Z(t),\\
& \rho = \sqrt{\xi^2 + s^2 + 2\xi s\cos\psi}.
\end{split}
\end{equation}
Since \eqref{eq:phi_t} depends explicitly on $\xi$ and $\psi$ only, it is convenient to perform calculations in the moving reference frame: $(\rho,\chi)\mapsto (\xi,\psi)$, see Fig.~\ref{fig:mobile}. In this frame, every integral over the domain $|z|<R$ can be calculated as
\begin{equation} \label{eq:dr-mov1}
\int_0^{2\pi} \mathrm{d}\chi \int_0^1 \rho \mathrm{d}\rho  f(\bullet) =  \int_0^{2\pi} \mathrm{d}\psi \int_0^{\sigma(\psi)} \xi \mathrm{d}\xi  f(\bullet),
\end{equation}
where $\sigma(\psi)$ is given by expressions
\begin{equation} \label{eq:sigma}
\begin{split}
& \sigma(\psi) = -s\cos\psi + \sqrt{1-s^2\sin^2\psi}.
\end{split}
\end{equation}
The gyroscopic term $G_1$ can be written as follows:
\begin{equation} \label{eq:G1}
\begin{split}
	G_1 &=\frac{p\left(1+R^2\right)}{\pi} \left( \dot{s} I_1  -  \dot{\Phi} I_2\right),
\end{split}
\end{equation}

where $I_1$ and $I_2$ can be calculated for the case of a small vortex displacement, $s \ll 1$:
\begin{equation} \label{eq:I1I2}
\begin{split}
I_1 &= \!\!\int_0^{2\pi} \!\! \mathrm{d}\psi  \!\!\int_0^{\sigma(\psi)} \!\! \frac{ \phi_{\text{s}}\xi\mathrm{d}\xi}{\left[1+ R^2\rho^2(\xi) \right]^2} \approx 0,\\
I_2 &= \!\!\int_0^{2\pi}\!\! \mathrm{d}\psi \!\!\int_0^{\sigma(\psi)} \!\! \frac{ \phi_{\Phi}\xi\mathrm{d}\xi}{\left[1+ R^2\rho^2(\xi) \right]^2} \approx-\frac{\pi s}{1+R^2s^2}.
\end{split}
\end{equation}
In fact, while the result obtained in Eq.~(\ref{eq:I1I2}) is valid for any vortex displacement, the integrals $I_1$ and $I_2$ can be calculated analytically only for the case of small vortex displacements.

The gyroscopic term $G_2$, see \eqref{eq:G1&G2} can be calculated as follows. Since the out--of--surface vortex structure $\cos\theta^{\textsc{iva}}$ is localized with a typical size of the vortex core and taking into account that $\dot\phi$ takes the form of Eq.~(\ref{eq:phi_t}), the second gyroscopic term $G_2=0$. Finally, the gyroscopic term takes the form \eqref{eq:G-eff}.

%%%%%%%%%%%%%%%%%%%%%%%%%%%%%%%%%%%%%%%%%%%%%%%%%%
%%%	New Section
%%%%%%%%%%%%%%%%%%%%%%%%%%%%%%%%%%%%%%%%%%%%%%%%%%

\section{Calculation of the effective dissipative function}
\label{sec:Fdiss}

In the same way as above, we assume that the system size $L$ is much larger than the exchange length $\ell$. Thus, the main contribution to the dissipative function is due to the in--surface vortex distribution,
\begin{equation} \label{eq:Fdiss-ap1}
\begin{split}
F &= \frac{\alpha}{V} \int \mathrm{d} \vec{r} \left[ \left(\dot{\theta}^{\textsc{iva}} \right)^2 + \sin^2\left(\theta^{\textsc{iva}}\right) \left(\dot{\phi}^{\textsc{iva}}\right)^2 \right]\\
& \approx \frac{\alpha (1+R^2)}{\pi} \int_0^{2\pi}\mathrm{d}\psi \!\! \int_{\lambda}^{\sigma(\psi)} \!\! \frac{\xi \mathrm{d}\xi}{\left[1+R^2\rho^2(\xi)\right]^2}\left(\dot{\phi}^{\textsc{iva}}\right)^2.
\end{split}
\end{equation}
Here, we changed variables under the integral using rules \eqref{eq:dr} and \eqref{eq:dr-mov1}. By incorporating the image--vortex Ansatz for the calculation of $\dot{\phi}^{\textsc{iva}}$ in the same manner as above in \eqref{eq:phi_t}, we rewrite the effective dissipative function as follows
\begin{equation} \label{eq:Fdiss-ap2}
\begin{split}
F \approx &2\alpha (1+R^2) \Biggl[ {\dot{s}}^2 \Bigl\langle\Upsilon(\phi_{\text{s}}^2) \Bigr\rangle + \dot{\Phi}^2 \Bigl\langle\Upsilon(\phi_{\Phi}^2) \Bigr\rangle \\
& -  \dot{s} \dot{\Phi} \Bigl\langle\Upsilon(2\phi_{\text{s}}\phi_{\Phi}) \Bigr\rangle \Biggr].
\end{split}
\end{equation}
Here, the function $\Upsilon(f(\psi,\xi))$ has the following form
\begin{equation} \label{eq:Upsilon}
\Upsilon\left(f(\psi,\xi)\right) = \int_{\lambda}^{\sigma(\psi)}\frac{f(\psi,\xi)\mathrm{\xi d}\xi}{ \left[1+R^2\rho^2(\xi)\right]^2},
\end{equation}
and the averaging means $\Bigl\langle f(\psi) \Bigr\rangle = (1/2\pi) \int\limits_0^{2\pi}f(\psi) \mathrm{d}\psi$. The functions $\Upsilon$ in Eq.~(\ref{eq:Fdiss-ap2}) can be calculated for the case of small vortex displacements, $s\ll1$, under the condition $\ln\left(1/\lambda\right)\gg1$,
\begin{equation} \label{eq:Upsilon2}
\begin{split}
	& \Upsilon(\phi_{\text{s}}^2) \approx \Upsilon_0 \sin^2\psi,\;\;\;\;\;\;\;\;\;\;\;\;\; \Upsilon(\phi_{\Phi}^2) \approx\Upsilon_0 s^2 \cos^2\psi, \\
	& \Upsilon(2 \phi_{\text{s}}\phi_{\Phi}) \approx \Upsilon_0 s \sin(2\psi),\;\;\; \Upsilon_0 = \frac{\ln(1/\lambda)}{\left(1+R^2s^2\right)^2}.
\end{split}
\end{equation}
Using \eqref{eq:Upsilon2}, we calculate the effective dissipative function \eqref{eq:Fdiss-ap1}, which results in \eqref{eq:F-eff}.

%%%%%%%%%%%%%%%%%%%%%%%%%%%%%%%%%%%%%%%%%%%%%%%%%%
%%%	New Section
%%%%%%%%%%%%%%%%%%%%%%%%%%%%%%%%%%%%%%%%%%%%%%%%%%

\section{Calculation of the magnetostatic energy}
\label{sec:Magnetostatic-app}
To calculate the magnetostatic energy as a function of the vortex position, we consider the contribution of the volume magnetostatic charges $\lambda_{\text{ms}} = -\vec{\nabla}\cdot \vec{m}$ only. For this purpose, we map the spherical surface coordinate to a complex plane: $\vec{r}\mapsto w\mapsto z$, see Fig.~\ref{fig:mapping}. By neglecting the out--of--surface contribution of the vortex structure and relying on the image vortex Ansatz \eqref{eq:iva}, we calculate the magnetostatic charges as follows:
\begin{equation} \label{eq:lambda-app}
	\begin{split}
		&\lambda_{\text{ms}}(\vec r)=\frac{\mathfrak{C} s(1+R^2)}{R r} \Lambda(\rho,s,\chi-\Phi),\\
		&\Lambda(\rho,s,\alpha) \\
		&=\frac{\rho \sin\alpha}{\sqrt{\rho^2 + s^2 - 2\rho s \cos\alpha} \sqrt{1 + \rho^2 s^2 -2\rho s \cos\alpha}}.\!\!\!
	\end{split}
\end{equation}

Then the magnetostatic vortex energy reads
\begin{equation} \label{eq:Ems-via-C-app}
\begin{split}
\mathscr{E}_{\text{ms}} &= \frac{1}{8\pi V}\!\!\int\!\!\mathrm{d}\vec r \!\!\int \!\! \mathrm{d}\vec r'\frac{\lambda_{\text{ms}}(\vec r)\lambda_{\text{ms}}(\vec r')}{|\vec r-\vec r'|} \\
&= \varepsilon s^2 \mathcal{C}(\nu,s,\vartheta_0) = \varepsilon \frac{1-\cos\Theta}{1-\cos\vartheta_0}\mathcal{	C}(\nu,s,\vartheta_0)
\end{split}
\end{equation}

with the stiffness parameter $\mathcal{C}(\nu,s,R)$ given by the expression:
\begin{widetext}
	\begin{equation} \label{eq:C}
	\begin{split}
	\mathcal{C}(\nu,s,\vartheta_0) &=
	\frac{3 (\nu  (\nu +2)+2) R \left(R^2+1\right)^2 \left(R^2 s^2+1\right)}{\pi ^2 \nu ^2 (\nu +2) (\nu  (\nu +3)+3)}
	 \int_0^{1+\nu} \mathrm{d} x  \int_0^{1+\nu} \mathrm{d} x' \int_0^1 \mathrm{d}\rho \int_0^1 \mathrm{d}\rho' \int_0^{2\pi}  \mathrm{d}\chi \int_0^{2\pi}  \mathrm{d}\chi' W_{\text{ms}},\\
	W_{\text{ms}} =&
	\frac{\rho  \rho '}{\left(1+\rho^2 R^2\right)^2 \left(1+R^2 \rho '^2\right)^2}
	\frac{x x' \Lambda(\rho,s,\chi)\Lambda(\rho',s,\chi')}{\sqrt{x^2+x'^2-\frac{2 x x' \left(4 \rho ^2 \rho '^2 \cos \left(\chi -\chi '\right)+\left(\rho ^2-1\right) \left(\rho '^2-1\right)\right)}{\left(\rho ^2+1\right) \left(\rho '^2+1\right)}}},\; R=\tan\frac{\vartheta_0}{2}.
	\end{split}
	\end{equation}	
	
	Under assumption of small vortex displacements ($s \ll 1$), $\Lambda(\rho,s \to0,\alpha) = \sin\alpha$. To calculate $\mathcal{C}(\nu,0,R)$, we expand $1/|\vec{r} - \vec{r}'|$ in associated Legendre polynomials and rewrite $\mathcal{C}(\nu,0,\vartheta_0)$ as follows:
	\begin{equation} \label{eq:Ems&C}
	\begin{split}
	&\mathcal{C}(\nu,0,\vartheta_0) = \frac{1}{\sin\vartheta_0 \left(1-\cos\vartheta_0\right)} \sum_{l=1}^\infty \frac{f_l(\nu)I_l^2(\vartheta_0)}{l(l+1)},\qquad I_l(\vartheta_0) = \int_{\cos\vartheta_0}^1\!\!\! P_l^1(z) \mathrm{d}z,\\
	f_l(\nu) &= \begin{cases}
	\dfrac{(\nu  (\nu +2)+2) (\nu  (\nu  (\nu +3)+3)-3 \log (\nu +1))}{3 \nu ^2 (\nu +2) (\nu  (\nu +3)+3)}, & \text{when $l=1$},\\
	\dfrac{(\nu  (\nu +2)+2) (\nu +1)^{-l} \left(((l-1) \nu  (\nu  (\nu +3)+3)-3) (\nu +1)^l+3 (\nu +1)\right)}{\left(l^2+l-2\right) \nu ^2 (\nu +2) (\nu  (\nu +3)+3)},& \text{when $l\neq 1$}.
	\end{cases}
	\end{split}
	\end{equation}
	
	In the case of a very thin cap $\nu \to 0$, we obtain $f_l(\nu \to 0) = \dfrac{1}{2}$ and $\mathcal{C}(\vartheta_0)$, which results in (\ref{eq:C4nu0}).
\end{widetext}
%
%%%%%%%%%%%%%%%%%%%%%%%%%%%%%%%%%%%%%%%%%%%%%%%%%%
%%%		Bibliography
%%%%%%%%%%%%%%%%%%%%%%%%%%%%%%%%%%%%%%%%%%%%%%%%%%
%\bibliography{soliton}

\end{document}